\documentclass[aps,prb,reprint,groupedaddress]{revtex4-1}

\usepackage{graphicx}
\usepackage{amssymb}

\begin{document}

%\preprint{}

\title{Electronic entropy in Green's-function calculations
       at finite temperatures}


\author{I. Turek}
\email[]{turek@ipm.cz}
%\thanks{}
\affiliation{Institute of Physics of Materials,
Academy of Sciences of the Czech Republic,
\v{Z}i\v{z}kova 22, CZ-616 62 Brno, Czech Republic}

\author{J. Kudrnovsk\'y}
\email[]{kudrnov@fzu.cz}
\affiliation{Institute of Physics, 
Academy of Sciences of the Czech Republic,
Na Slovance 2, CZ-182 21 Praha 8, Czech Republic}

\author{V. Drchal}
\email[]{drchal@fzu.cz}
\affiliation{Institute of Physics, 
Academy of Sciences of the Czech Republic,
Na Slovance 2, CZ-182 21 Praha 8, Czech Republic}


\date{\today}

\begin{abstract}
We revise critically existing approaches to evaluation of
thermodynamic potentials within the Green's function calculations
at finite electronic temperatures.
We focus on the entropy and show that usual technical problems 
related to the multivalued nature of the complex logarithm can be
overcome.
This results in a simple expression for the electronic entropy,
which does not require any contour integration in the complex
energy plane.
Properties of the developed formalism are discussed and its
illustrating applications to selected model systems and to bcc
iron with disordered local magnetic moments are presented as well. 
\end{abstract}

%\pacs{72.10.Bg, 72.25.Rb, 75.78.-n}
%\keywords{}

\maketitle

%\subsection{}
%\subsubsection{}

\section{Introduction\label{s_intr}}

The effect of finite temperatures on structure and properties
of metallic systems is of general importance for the whole
solid-state physics.
On the theoretical side, this fact can be illustrated by existing
systematic studies of alloy phase stability \cite{r_1991_fd} and
of spin fluctuations in itinerant magnets \cite{r_1985_tm}.
More recently, a number of \emph{ab initio} theoretical studies
have appeared dealing, e.g., with magnetic anisotropy in layered
systems \cite{r_2007_bws}, transport properties and damping of
magnetization dynamics \cite{r_2015_emc}, interplay of magnetism
and thermal lattice expansion \cite{r_2017_dlc}, or phase
stability and magnetism of iron under Earth's core conditions
\cite{r_2017_blf, r_2013_rbs}.
Reliable theoretical approaches to equilibrium properties at
finite temperatures have to provide not only the total energy of
the studied systems, but also other thermodynamic potentials and
quantities, in particular the free energy, the grand canonical
potential, and the entropy.

This task represents a challenge for material-specific theory
for several reasons.
First, all relevant temperature-induced excitations (phonons,
magnons, electrons) should be taken into account in a consistent
manner.
Second, the elevated temperatures lead to structure defects, such
as vacancies, impurities, antisite atoms, chemical disorder, etc.,
which cannot be neglected especially in multicomponent systems with
several sublattices.
Third, the presence of excitations and structure defects violates
the perfect translation invariance, so that standard techniques of 
\emph{ab initio} electron theory of solids, employing the well-known
Bloch theorem, are of limited applicability.
Systems with broken translation invariance are often treated by
means of Green's-function techniques \cite{r_1990_pw, r_1992_ag,
r_1997_tdk}.

The use of the Green's functions in first-principles calculations
with finite temperatures (within the general density-functional
theory \cite{r_1965_ndm}) was worked out by various authors a long
time ago \cite{r_1995_wlz, r_1997_nz, r_2005_rz}.
As a rule, all of these schemes employ complex energy variables
and integrations over contours in the complex plane which increases
substantially the computational efficiency \cite{r_1982_zdd, 
r_1982_wfl}.
The essence of this advantage lies in the analytic and smooth
behavior of the Green's function (resolvent of the effective
one-particle Hamiltonian) for energy arguments lying deeply in the
complex plane, in contrast to non-analytic and sharp spectral
features often encountered on the real energy axis. 

The developed finite-temperature Green's-function techniques
\cite{r_1995_wlz, r_1997_nz} work surprisingly well for obtaining
self-consistent electron densities and total energies; however, an
unpleasant drawback appears in evaluation of the
entropy \cite{r_1997_nz}, the free energy, or the grand canonical
potential \cite{r_1995_wlz}.
The origin of this feature can be traced back to the branch cut
of the complex logarithmic function which enters the expression
for entropy
(to be specific, we have in mind entropy due to the particle-hole
excitations as described by the Fermi-Dirac occupation function).
This branch cut prevents flexible deformations of complex
integration contours, which calls for alternative means in order
to obtain reliable results \cite{r_1997_nz}.
One way to circumvent this problem is the use of an expression for
the grand canonical potential which does not contain the logarithmic
function explicitly, see Eq.~(33) in Ref.~\onlinecite{r_1995_wlz}.
However, this alternative approach requires the integrated density
of states as a function of the complex argument; since this function
is obtained typically from the logarithm of determinant of a secular
matrix, the problem of the multivalued complex logarithm does not
seem to be removed completely from the formalism.

The purpose of the present paper is to derive another expression for
entropy in the Green's-function calculations which is not affected
by the above-mentioned problems due to the ambiguity of complex
logarithm. 
It turns out that the derived formula is even simpler than existing
expressions for other quantities; in particular, it does not contain
any explicit contour integration.
The new expression for entropy and accompanying expressions for
electron densities and other quantities are implemented in the
first-principles tight-binding (TB) linear muffin-tin orbital (LMTO)
method \cite{r_1984_aj, r_1997_tdk}; numerical tests of accuracy for
selected model and realistic systems are presented as well.

\section{Theoretical formalism\label{s_theo}}

\subsection{Electron densities and energies\label{ss_ede}}

The majority of usual quantities in \emph{ab initio} electron
theory of solids, such as the valence contribution to electron
densities in the real space or to the sum of occupied
one-particle eigenvalues, can be written as integrals over the
real energy axis
\begin{equation}
Q = \int_{-\infty}^{+\infty} D(E) f(E) \mathrm{d}E ,
\label{eq_qdef}
\end{equation}
where the function $D(E)$ is closely related to a projected or
total density of states (DOS) and $E$ is a real energy variable.
The function $f(E)$ denotes the Fermi-Dirac distribution
$f(E) = \{ 1 + \exp [\beta(E-\mu)] \}^{-1}$,
where $\mu$ is the chemical potential and $\beta$ refers
to the reciprocal value of the finite temperature $T$ as
$\beta = (k_\mathrm{B} T)^{-1}$, where $k_\mathrm{B}$ is the
Boltzmann constant.
In Green's-function techniques, the function $D(E)$ can be
written as
\begin{equation}
D(E) = \lim_{\varepsilon \to 0^+} \, 
\frac{i}{2\pi} \left[ \Gamma(E+i\varepsilon) - 
\Gamma(E-i\varepsilon) \right] , 
\label{eq_ddef}
\end{equation}
where the function $\Gamma(z)$ of a complex variable $z$ 
depends linearly on the resolvent $G(z) = (z - H)^{-1}$ of
the underlying one-particle Hamiltonian $H$.
Note that $H$ and $G(z)$ are operators (matrices), whereas
$D(E)$ and $\Gamma(z)$ are usual real and complex functions,
respectively.
We assume that the function $\Gamma(z)$ is analytic everywhere
in the complex plane with exception of real energies belonging
to the spectrum of $H$.

\begin{figure}[ht]
\begin{center}
\includegraphics[width=0.80\columnwidth]{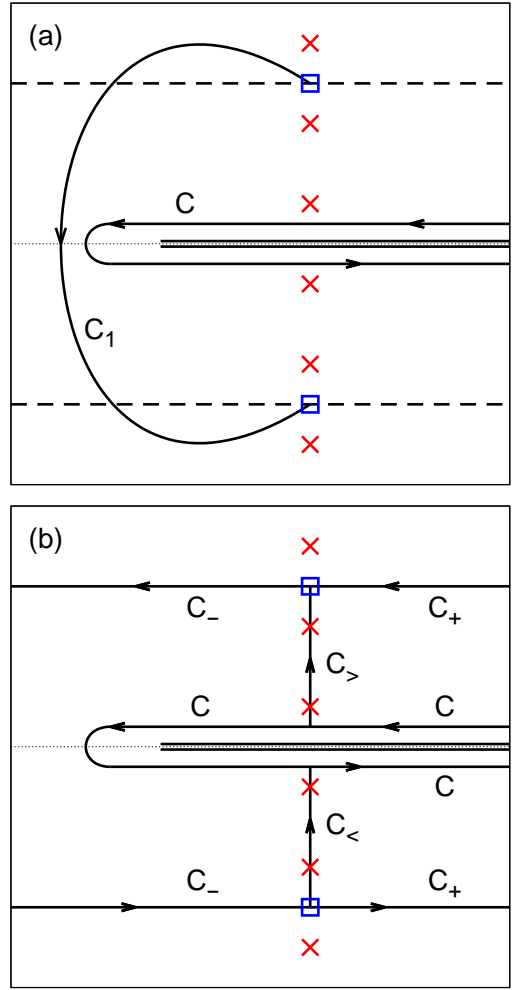}
\end{center}
\caption{
(a) The integration contours $C$ and $C_1$ in Eq.~(\ref{eq_qcint})
and Eq.~(\ref{eq_q1}), respectively. 
The horizontal double line marks the valence part of the
spectrum of Hamiltonian $H$, 
the crosses denote the Matsubara points $z_k$ and the squares
denote the points $Z_M$ and $Z^\ast_M$ for $M=2$.
(b) The integration contours $C$, $C_+$, $C_-$, $C_>$, and
$C_<$ for entropy calculation.
\label{f_cint}}
\end{figure}

The standard transformation of Eq.~(\ref{eq_qdef}) into a
complex integral starts from the form
\begin{equation}
Q = \frac{1}{2\pi i} \int_C \Gamma(z) f(z) \mathrm{d}z ,
\label{eq_qcint} 
\end{equation}
where the complex contour $C$ is drawn around the valence part
of the spectrum of $H$ as shown in Fig.~\ref{f_cint}a.
The modification of Eq.~(\ref{eq_qcint}) rests on the well-known
properties of the functions $\Gamma(z)$ and $f(z)$, such as their
analyticity, the periodicity of $f(z)$ with an imaginary period
$2\pi i \beta^{-1}$, the existence of simple poles of $f(z)$
at the Matsubara points $z_k = \mu + \pi i \beta^{-1} k$, where
$k$ runs over all odd integers,
an exponential decay of $f(z)$ for $\Re (z) \to +\infty$, and 
an exponential decay of $[f(z)-1]$ for $\Re (z) \to -\infty$, 
where $\Re (z)$ denotes the real part of $z$.

The final form of $Q$, described in the literature \cite{r_1995_wlz},
employs a point $Z_M = \mu + 2\pi i \beta^{-1} M$, where $M$ is a
positive integer; note that there are exactly $M$ Matsubara points
$z_k$ lying inside the segment $( \mu, Z_M )$. 
This yields:
\begin{equation}
Q = Q_1 + Q_2 + Q_3 ,
\label{eq_qfin}
\end{equation}
with the individual terms given by 
\begin{equation}
Q_1 = \frac{1}{2\pi i} \int_{C_1} \Gamma(z) \mathrm{d}z ,
\label{eq_q1}
\end{equation}
where the contour $C_1$ starts at $Z_M$ and ends at $Z^\ast_M$,
see Fig.~\ref{f_cint}a,
\begin{equation}
Q_2 = \beta^{-1} \sum_{|k| < 2M}^\mathrm{odd} \Gamma(z_k) ,
\label{eq_q2}
\end{equation}
where the sum runs over all Matsubara points $z_k$ between $Z_M$
and $Z^\ast_M$, and
\begin{eqnarray}
Q_3 & = &
\frac{i}{2\pi} \int_{-\infty}^{+\infty} \Gamma(Z_M + \xi) \,
\frac{\mathrm{sign}(\xi)\, \mathrm{d}\xi}{1 + \exp(\beta |\xi| )}
\nonumber\\
 & & {} - 
\frac{i}{2\pi} \int_{-\infty}^{+\infty} \Gamma(Z^\ast_M + \xi) \,
\frac{\mathrm{sign}(\xi)\, \mathrm{d}\xi}{1 + \exp(\beta |\xi| )} ,
\label{eq_q3}
\end{eqnarray}
where the integrations over a real variable $\xi$ correspond to 
complex integrals along the horizontal dashed lines in
Fig.~\ref{f_cint}a.
Note that the contributions over $\xi > 0$ in Eq.~(\ref{eq_q3})
refer to integrations of $f(z) \Gamma(z)$, whereas those over
$\xi < 0$ refer to integrations of $[f(z)-1] \Gamma(z)$. 
By employing the rule $\Gamma(z^\ast)=\Gamma^\ast(z)$, the
integrations in $Q_1$ and $Q_3$ and the summation in $Q_2$ can be
performed with arguments $z$ of $\Gamma(z)$ lying only in the upper
(or lower) complex half-plane, see Ref.~\onlinecite{r_1995_wlz} for
more details.

Numerically, the integral in $Q_1$ (\ref{eq_q1}) can be evaluated
by using a finite number of nodes along the path $C_1$ and
corresponding weights,
and the integrals in $Q_3$ (\ref{eq_q3}) can be obtained from a
finite number of terms of the Taylor expansion of $\Gamma(z)$ at
the point $Z_M$.
The coefficients of this expansion are obtained from the values
of $\Gamma(z)$ in a few points $z$ lying in the neighborhood of
$Z_M$ and the remaining integrals can be reduced (for odd positive
integers $j$) to
\begin{equation}
R_j = \int_{-\infty}^{+\infty} 
\frac{|u|^j \, \mathrm{d}u}{ 1 + \mathrm{e}^{|u|} }
= 2 \, j! \, (1 - 2^{-j}) \, \zeta(j+1) ,
\label{eq_rj}
\end{equation}
where $\zeta(a) = \sum_{n=1}^\infty n^{-a}$ is the Riemann's
zeta-function (for $a > 1$).

\subsection{Entropy\label{ss_entr}}

The entropy corresponding to a positive temperature and the
one-particle Hamiltonian $H$ is given by
\begin{equation}
S = k_\mathrm{B}
\int_{-\infty}^{+\infty} D(E) \sigma(E) \mathrm{d}E ,
\label{eq_sdef}
\end{equation}
where $D(E)$ is the total DOS of the system related to
the Green's function (resolvent) by Eq.~(\ref{eq_ddef})
with $\Gamma(z) = \mathrm{Tr} \{ G(z) \}$,
and where the function $\sigma(E)$ is defined as
\begin{equation}
\sigma(E) = - f(E) \ln [f(E)] - [1-f(E)] \ln [1-f(E)] .
\label{eq_sigma}
\end{equation}
Since complex logarithm is a multivalued function, the real
function $\sigma(E)$ can be directly continued analytically into
its complex counterpart $\sigma(z)$ only in a stripe around the
real energy axis, namely, for $| \Im(z) | < \pi\beta^{-1}$,
where $\Im(z)$ denotes the imaginary part of $z$.
[Here and below, we assume the branch cut of $\ln(w)$ along the
real negative half-axis, $w < 0$.]
For present purposes, let us define continuations $\sigma_-(z)$
and $\sigma_+(z)$ of $\sigma(E)$ that are analytic in the entire
half-planes $\Re(z) < \mu$ and $\Re(z) > \mu$, respectively:
\begin{equation}
\sigma_\pm(z) = \frac{\pm t}{ 1 + \mathrm{e}^{\pm t} } 
  + \ln ( 1 + \mathrm{e}^{\mp t} ) ,  \ \quad 
  t = \beta (z-\mu) ,
\label{eq_sigmapm}
\end{equation}
and let us discuss briefly their properties.
First, the function $\sigma_+(z)$ decays exponentially for
$\Re(z) \to +\infty$ and
the function $\sigma_-(z)$ decays exponentially for
$\Re(z) \to -\infty$. 
Second, it can be shown that $\sigma_+(z)$ and $\sigma_-(z)$
possess the same leading term of their singular behavior near the
Matsubara points $z_k$: 
\begin{equation}
\sigma_\pm(z) = (\mu - z_k) (z - z_k)^{-1} + \dots ,
\label{eq_singsig}
\end{equation}
where the omitted term includes a regular part and a weak
(logarithmic) singularity.
Third, along the vertical line $z = \mu + i\eta$ ($\eta$ is real),
the limits of $\sigma_\pm(\mu\pm\varepsilon + i\eta)$ for 
$\varepsilon \to 0^+$ can be considered; their difference equals
\begin{eqnarray}
\tau(\mu + i\eta) & = & \lim_{\varepsilon \to 0^+}
 [ \sigma_+(\mu + \varepsilon + i\eta) - 
   \sigma_-(\mu - \varepsilon + i\eta) ]
\nonumber\\
 & = & 2\pi i \, [[ \beta\eta / (2\pi) ]] ,
\label{eq_tau}
\end{eqnarray}
where $[[ u ]]$ denotes the integer nearest to the real
quantity $u$.
This result means that $\tau(\mu + i\eta)$ is
a piecewise constant function of $\eta$ with discontinuities
at $\eta = \pi \beta^{-1} k$, where $k$ is an odd integer
(i.e., whenever $\mu + i\eta = z_k$). 
The last property reflects the fact that the derivatives of 
$\sigma_+(z)$ and $\sigma_-(z)$ coincide mutually in the whole
complex plane except at the Matsubara points $z_k$, where
second-order poles of both derivatives are located.
Fourth, along horizontal lines $z = E +  2\pi i \beta^{-1} m$, 
where $m$ is an integer, it holds
\begin{equation}
\sigma_\pm(E + 2\pi i \beta^{-1} m) = \sigma(E) \pm 
\frac{2\pi i m}{1 + \exp[ \pm \beta (E-\mu) ] } \, ,
\label{eq_sigx}
\end{equation}
proving explicitly that the functions $\sigma_\pm(z)$ are not
periodic with the period $2\pi i \beta^{-1}$, in contrast to
the function $f(z)$.

The expression for the entropy (\ref{eq_sdef}) can be written
as a complex integral
\begin{equation}
\frac{S}{k_\mathrm{B}} = \frac{1}{2\pi i}
\int_C \Gamma(z) \sigma(z) \mathrm{d}z 
\label{eq_soldci}
\end{equation}
along the same path $C$ as in Eq.~(\ref{eq_qcint}).
The deformation of the contour $C$ has to be performed separately
on both sides of the vertical line $\Re(z) = \mu$,
see Fig.~\ref{f_cint}b.
This leads to the form
\begin{equation}
 S = S_+ + S_- + S_> + S_< 
\label{eq_ssum}
\end{equation}
with the individual terms given by
\begin{eqnarray}
\frac{S_\gtrless}{k_\mathrm{B}} & = & \frac{i}{2\pi}
\int_{C_\gtrless} \Gamma(z) \tau(z) \mathrm{d}z  +  
\sum_{|k| < 2M}^{k \gtrless 0, \mathrm{odd}} (z_k - \mu)
\Gamma(z_k) ,
\nonumber\\
\frac{S_\pm}{k_\mathrm{B}} & = & \frac{1}{2\pi i}
\int_{C_\pm} \Gamma(z) \sigma_\pm(z) \mathrm{d}z ,
\label{eq_snewci}
\end{eqnarray}
where the paths $C_+$, $C_-$, $C_>$, and $C_<$ are depicted in
Fig.~\ref{f_cint}b and where the functions $\sigma_\pm(z)$ 
(\ref{eq_sigmapm}), $\tau(z)$ (\ref{eq_tau}), and the singular 
behavior of $\sigma_\pm(z)$ (\ref{eq_singsig}) have been used.

The contribution $S_+$ is calculated from Eq.~(\ref{eq_sigx})
for $m = M$ and $m = -M$, and similarly for $S_-$.
This yields together
\begin{eqnarray}
\frac{S_+ + S_-}{k_\mathrm{B}} = 
 - \frac{1}{\pi} \int_{-\infty}^{+\infty} 
\Im [ \Gamma(Z_M + \xi) ] \, \sigma(\mu + \xi) \, \mathrm{d}\xi
 & &
\nonumber\\
 {} - 2M \int_{-\infty}^{+\infty} \Re [ \Gamma(Z_M + \xi) ] \, 
\frac{\mathrm{sign}(\xi)\, \mathrm{d}\xi}{1 + \exp(\beta|\xi|)}
\, . \quad
 & & 
\label{eq_splmi}
\end{eqnarray}
Numerically, the integrals in Eq.~(\ref{eq_splmi}) can again be
obtained from a finite number of terms of the Taylor expansion
of $\Gamma(z)$ at the point $Z_M$.
The encountered integrals are $R_j$ (\ref{eq_rj}) and
(for even non-negative integers $j$) 
\begin{eqnarray}
N_j & = & \int_{-\infty}^{+\infty} u^j \left[
\frac{\ln( 1 + \mathrm{e}^u ) }{1 + \mathrm{e}^u} +
\frac{\ln( 1 + \mathrm{e}^{-u} ) }{1 + \mathrm{e}^{-u}}
 \right] \mathrm{d}u
\nonumber\\
 & = & \frac{j+2}{j+1} \, R_{j+1} \, .
\label{eq_nj}
\end{eqnarray}

The evaluation of $S_>$ and $S_<$ in Eq.~(\ref{eq_snewci}) is
greatly simplified by the fact that the function $\tau(z)$ defined
along the paths $C_>$ and $C_<$ is piecewise constant, see
Eq.~(\ref{eq_tau}).
If we denote by $\Phi(z)$ a primitive function to $\Gamma(z)$,
so that $\mathrm{d}\Phi(z) / \mathrm{d}z = \Gamma(z)$, we get
\begin{eqnarray}
\frac{S_>}{k_\mathrm{B}} & = &
 \sum_{0 < k < 2M}^\mathrm{odd} \Phi(z_k) - M \Phi(Z_M)
\nonumber\\
 & & 
{} + \sum_{0 < k < 2M}^\mathrm{odd} (z_k - \mu) \Gamma(z_k) ,
\label{eq_sgr}
\end{eqnarray}
and similarly for $S_<$.
The primitive function can be chosen to satisfy the rule
$\Phi(z^\ast) = \Phi^\ast(z)$; in such a case, one obtains
\begin{eqnarray}
\frac{S_> + S_<}{k_\mathrm{B}} & = & 
2 \Re \Bigg\{ \sum_{0 < k < 2M}^\mathrm{odd}
 \left[ \Phi(z_k) + (z_k - \mu) \Gamma(z_k) \right] \Bigg.
\nonumber\\
 & &
\Bigg. {} - M \Phi(Z_M) \Bigg\} .
\label{eq_sgrle}
\end{eqnarray}
This result means that only the values of $\Gamma(z)$ and
$\Phi(z)$ at the point $Z_M$ and at the $M$ lowest Matsubara
points $z_k$ in the upper half-plane are needed.
In practice, the function $\Phi(z)$ is often constructed as
$\Phi(z) = \ln(| z - H |)$, where $| z - H |$ denotes the
determinant of the secular matrix. 
Note that the ambiguity of the imaginary part of logarithm does
not affect the obtained result (\ref{eq_sgrle}), which depends only
on the unambiguous real part of the logarithmic function.

Let us conclude this section by a few comments to the derived final
expression for the entropy $S$, given by the sum of
Eq.~(\ref{eq_splmi}) and Eq.~(\ref{eq_sgrle}). 
First, this final result includes no explicit contour integral,
so that it is simpler than the final result for quantities $Q$
treated in Section~\ref{ss_ede}.
This feature can be ascribed to different asymptotic behavior of
the functions $\sigma(E)$ and $f(E)$:
the former decays exponentially for $E \to \pm \infty$, whereas
the latter decays only for $E \to +\infty$, but it approaches unity
for $E \to -\infty$. 
Second, the primitive function $\Phi(z)$ in Eq.~(\ref{eq_sgrle}),
closely related to the integrated DOS, can be constructed from
the determinant of the secular matrix not only in a general theory
considered here, but also in the multiple-scattering 
Korringa-Kohn-Rostoker (KKR) theory \cite{r_1995_wlz}
or in the LMTO method \cite{r_1997_tdk, r_1996_dkp}.
Moreover, for substitutionally disordered systems treated in the
coherent potential approximation (CPA), proper configuration
averages of the integrated DOS are available in the literature
\cite{r_1992_ag, r_2000_tkd}.
All these expressions for the primitive function $\Phi(z)$ contain
the logarithmic function, but the resulting right-hand side of
Eq.~(\ref{eq_sgrle}) is defined unambiguously again.
Finally, the additional numerical effort to calculate the entropy
is negligible as compared to other computations, since the
complex energy arguments involved (which comprise the $M$ lowest
Matsubara points $z_k$, the point $Z_M$, and a few points near
$Z_M$) enter the self-consistent electron-density calculations
as well (see end of Section~\ref{ss_ede}).

\section{Numerical implementation\label{s_numimp}}

The developed formalism has been implemented on a model level
and in the self-consistent scalar-relativistic TB-LMTO method
\cite{r_1984_aj} in the atomic sphere approximation (ASA) and
the CPA \cite{r_1997_tdk}.
Since the numerical evaluation of the entropy follows closely that of
the electron densities and effective potentials, the previous
experience has been employed to great extent \cite{r_1995_wlz}.
The integration contour $C_1$ (Fig.~\ref{f_cint}a) was a part
of a circle with the center located on the real energy axis;
the contour integration was performed numerically by means of
14 nodes (distributed along the upper half of $C_1$) and
corresponding complex weights.
In the illustrating example presented in Section~\ref{s_resdis},
several thousands of $\mathbf{k}$ vectors were used for sampling
the irreducible part of the bcc Brillouin zone (BZ) for the first
Matsubara point $z_1$ (closest to the real chemical potential
$\mu$), while reduced numbers of $\mathbf{k}$ vectors were used
for the complex energy points more distant from $\mu$. 

The coefficients of the Taylor expansions of the functions
$\Gamma(z)$ at $z = Z_M$ were obtained numerically based on the
calculated values in a few points in the distance $\sim \beta^{-1}
= k_\mathrm{B} T$ from $Z_M$.
In the simplest models, these coefficients were also set their
exact values for the sake of comparison of the effect of both
alternatives on the resulting electronic entropy.
The degree $\nu$ of the Taylor polynomials was varied in the range
$1 \le \nu \le 8$; however, for practical applications with
temperatures not exceeding $\sim 1000$~K, polynomials with
$\nu \le 4$ seem to be sufficient in most cases.

\section{Results and discussion\label{s_resdis}}

We start the discussion of accuracy of the developed formalism for
electronic entropy with analysis of a simple model DOS corresponding
to an isolated eigenvalue $E_0$ which coincides with the chemical
potential ($E_0 = \mu$), so that $D(E) = \delta(E-\mu)$, which
yields $\Gamma(z) = (z-\mu)^{-1}$, and $\Phi(z) = \ln (z-\mu)$.
We assume that $\mu$ is independent of temperature $T$, which leads
to the exact entropy $S_\mathrm{x}$ that is $T$-independent as well,
$S_\mathrm{x} = k_\mathrm{B} \ln 2$.
A closer look at Eq.~(\ref{eq_splmi}) and Eq.~(\ref{eq_sgrle}) in
this case reveals that they also provide $T$-independent values of
the approximate entropy $S$, which thus depends only on two integers
$M$ (number of the Matsubara points) and $\nu$ (degree of the Taylor
expansion polynomial).

\begin{figure}[ht]
\begin{center}
\includegraphics[width=0.95\columnwidth]{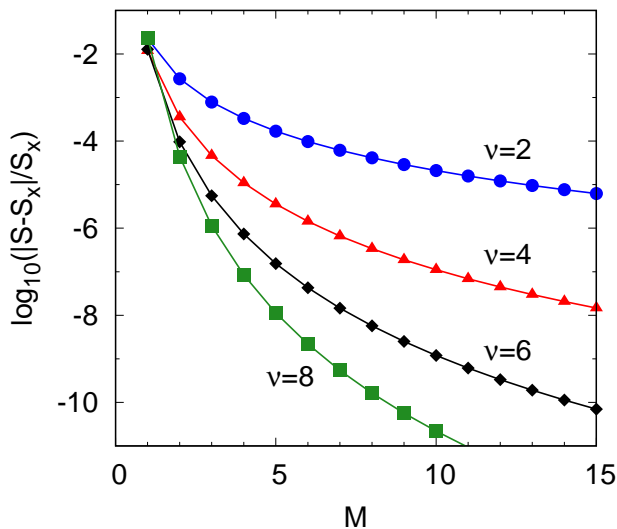}
\end{center}
\caption{
Relative deviations of the calculated entropy $S$ with respect to
the exact value $S_\mathrm{x} = k_\mathrm{B} \ln 2$ for the isolated
eigenvalue $E_0 = \mu$ and for different degrees $\nu$ of the Taylor
polynomial as functions of the number $M$ of Matsubara points.
Note the logarithmic scale on the vertical axis.
\label{f_isol}}
\end{figure}

The relative difference between $S$ and $S_\mathrm{x}$ is
displayed in Fig.~\ref{f_isol} as a function of $M$ for several
values of $\nu$.
One can see that the accuracy of the approximate scheme is quite
high, the only exceptions being the cases with very small $M$ or
$\nu$.
The studied model is very simple indeed; note however that an
isolated eigenvalue belongs to the strongest singularities
to be encountered in the electronic spectra.

\begin{figure}[ht]
\begin{center}
\includegraphics[width=0.95\columnwidth]{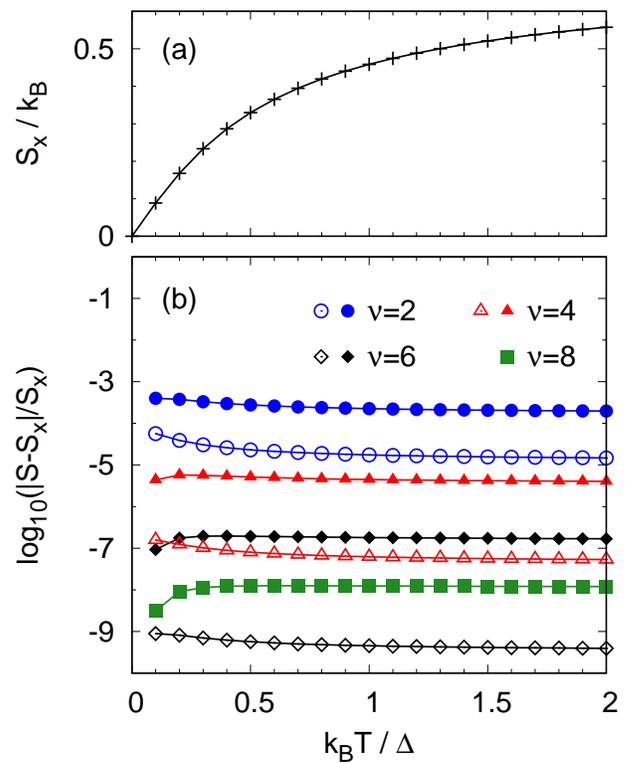}
\end{center}
\caption{
(a) The temperature dependence of the exact entropy $S_\mathrm{x}$
for a model Lorentzian DOS of width $\Delta$, see text for details.
(b) Relative deviations of the calculated entropy $S$ with respect
to $S_\mathrm{x}$ as functions of temperature for different degrees
$\nu$ of the Taylor polynomial and for two values of the number
$M$ of Matsubara points: $M=5$ (solid symbols) and $M=12$ (open
symbols).
\label{f_lor}}
\end{figure}

Another simple model is described by a Lorentzian DOS parametrized
by its center $E_0$ and width $\Delta$, so that
$D(E) \propto [ (E-E_0)^2 + \Delta^2 ]^{-1}$ and, consequently,
$\Gamma(z) = ( z - E_0 + i \Delta )^{-1}$ and
$\Phi(z) = \ln (z - E_0 + i \Delta)$ for $\Im(z) > 0$.
We have chosen a slight offset of the center $E_0$ with respect to
the chemical potential $\mu$, namely, $E_0 - \mu = 0.4 \Delta$, and
have assumed all parameters ($E_0$, $\Delta$, $\mu$) as
$T$-independent. 
The exact entropy $S_\mathrm{x}$, obtained by a highly accurate
real-energy numerical quadrature according to Eq.~(\ref{eq_sdef}),
is shown in Fig.~\ref{f_lor}a as a function of $T$.
The temperature dependence of the relative difference between the
approximate $S$ and the exact $S_\mathrm{x}$ is presented in 
Fig.~\ref{f_lor}b for two values of $M$ ($M = 5$ and $M = 12$) and
for several values of $\nu$.
One can see that the relative deviations are essentially independent 
of $T$, despite the pronounced increase of the entropy with
increasing temperature.
One can also observe an increase in the relative accuracy due to
higher values of $M$ and $\nu$, indicating that modest numbers $M$
and $\nu$ are sufficient for practical applications of the
developed formalism.

\begin{figure}[ht]
\begin{center}
\includegraphics[width=0.95\columnwidth]{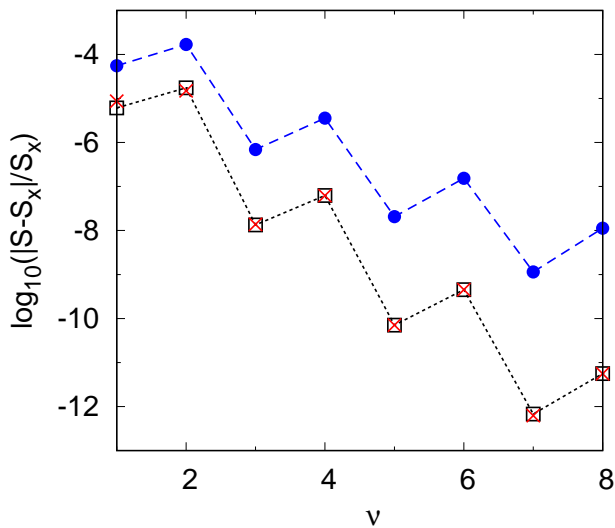}
\end{center}
\caption{
Relative deviations of the calculated entropy $S$ with respect to
the corresponding exact value $S_\mathrm{x}$ as functions of the
degree $\nu$ of the Taylor polynomial: for the isolated eigenvalue
with $M=5$ (solid circles) and for the Lorentzian DOS with
$k_\mathrm{B} T = \Delta$ and $M=12$ (open boxes).
The red crosses mark deviations obtained in the latter case with 
the Taylor expansion coefficients extracted by a numerical
procedure.
\label{f_nudep}}
\end{figure}

The origin of numerical inaccuracy in the present evaluation of
the entropy can easily be identified (disregarding the well-known
convergence issues with respect to the BZ sampling):
it is the treatment of integrations in Eq.~(\ref{eq_splmi}) by
using the Taylor expansion polynomials. 
Note that the same source of inaccuracy refers also to the quantity
$Q_3$ (\ref{eq_q3}).
There are two particular questions related to this point, namely,
(i) the effect of the finite degree $\nu$ of the Taylor polynomial,
and (ii) the role of the numerical procedure to extract the
coefficients of the polynomial.
The results found for the above simple models and presented in
Fig.~\ref{f_nudep} provide a partial answer to both questions.
First, it is seen that the increase in $\nu$ reduces in general
the relative deviations of entropy, but the obtained trends are not
strictly monotonic.
Second, the numerically obtained coefficients lead essentially to
the same values of $S$ as the exact coefficients (see the values
marked by red crosses and open boxes in Fig.~\ref{f_nudep}).

These facts represent undoubtedly positive features of the presented
formalism from the practical point of view; however, certain caution
is needed in attempts to increase the accuracy by using too high 
degrees $\nu$.
First, the convergence radius of the Taylor series of
$\Gamma(Z_M + \xi)$ is inevitably finite due to the branch cuts
(and possible poles) of $\Gamma(z)$ on the real energy axis.
This means that there is no strict convergence of $S$ to 
$S_\mathrm{x}$ for $\nu \to \infty$, at least for a fixed number
$M$ of the Matsubara points.
Second, the procedure to extract the Taylor coefficients from
several values of the function $\Gamma(z)$ in neighborhood of
$z = Z_M$ leads to a set of $\nu$ linear equations for $\nu$
unknown variables.
This linear problem (Vandermonde system) is ill-conditioned
\cite{r_1992_ptv} which can prevent its stable numerical solution
for large values of $\nu$.

\begin{table}
\caption{Calculated local magnetic moment $M_\mathrm{Fe}$
and electronic entropy $S$ (per atom) for three temperatures $T$
in the DLM state of bcc Fe.
The values of $M_\mathrm{Fe}$ in parenthesis are from
Ref.~\onlinecite{r_2013_rbs}.
\label{t_fems}}
\begin{ruledtabular}
\begin{tabular}{rcc}
$T$ (K) & $M_\mathrm{Fe}$ ($\mu_\mathrm{B}$) & $S / k_\mathrm{B}$ \\
\hline
  0     &    2.02 (1.96) &  0.00 \\
 2000   &    1.92 (1.85) &  0.89 \\
 4000   &    1.41 (1.30) &  2.09 \\
\end{tabular}
\end{ruledtabular}
\end{table}

\begin{figure}[ht]
\begin{center}
\includegraphics[width=0.90\columnwidth]{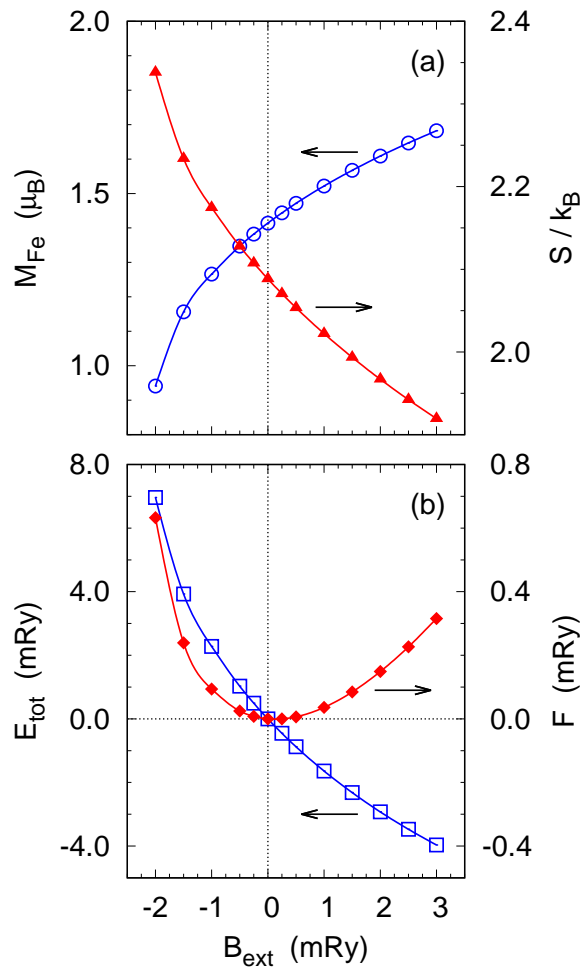}
\end{center}
\caption{
Calculated dependence of various quantities on the random external
magnetic field $B_\mathrm{ext}$ for bcc Fe in the DLM state at
$T = 4000$~K:
(a) the local magnetic moment $M_\mathrm{Fe}$ (open circles)
and the electronic entropy $S$ (solid triangles), and
(b) the total energy $E_\mathrm{tot}$ (open boxes)
and the free energy $F$ (solid diamonds).
The quantities $S$, $E_\mathrm{tot}$ and $F$ are given per one atom;
the vertical scales in panel (b) have been shifted to a common zero
at $B_\mathrm{ext} = 0$.
\label{f_fe}}
\end{figure}

As an illustrating application to a realistic system, we have
considered bcc iron in the disordered-local-moment (DLM) 
state \cite{r_1985_gps}.
This system at very high temperatures (up to $T = 6000$~K) and
under strong pressures attracts ongoing interest in the context
of physical properties of the Earth's core \cite{r_2017_blf,
r_2013_rbs, r_2016_pa, r_2017_dkw}.
Here we focus only on the effect of elevated temperatures and
treat thus bcc iron of a density corresponding to ambient
conditions (Wigner-Seitz radius $s = 2.65$~a.u.)
within the local spin-density approximation with the local
exchange-correlation potential parametrized according to
Ref.~\onlinecite{r_1980_vwn}.
The valence $spdf$-basis was used in the TB-LMTO-ASA method and
the CPA; the degree of the Taylor expansion polynomial was set
to $\nu = 6$.
For the very high temperatures considered here, small numbers
of the Matsubara points were sufficient: we set $M = 8$ for 
$T = 2000$~K and $M = 4$ for $T = 4000$~K.
Calculated values of the self-consistent local magnetic
moments $M_\mathrm{Fe}$ and of the electronic entropy $S$ (per
atom) are shown in Table~\ref{t_fems} for three selected
temperatures.
One can see a sizeable reduction of the local moment due to
increasing temperature as expected; the values of $M_\mathrm{Fe}$
in this work are slightly higher than the values reported in
Ref.~\onlinecite{r_2013_rbs}.
The electronic entropy increases with increasing temperature,
which is another expected trend of finite-temperature behavior.

In order to make better assessment of internal consistence of
the formulated entropy $S$ with other electronic quantities, such
as the electron densities, local magnetic moments $M_\mathrm{Fe}$
and total energies $E_\mathrm{tot}$, we have studied these
quantities as functions of a randomly oriented external magnetic
field $B_\mathrm{ext}$ coupled to the electron spin.
[For simplicity, the quantity $B_\mathrm{ext}$ includes the
Bohr magneton $\mu_\mathrm{B}$, so that $\pm B_\mathrm{ext}$
describes the spin-dependent shift added to the spin-polarized
exchange-correlation potential.]
The application of this random external field is equivalent to
an effect of the constraint in the fixed-spin-moment method applied
to the DLM state \cite{r_2017_dkw}.
Various calculated quantities for the case of $T = 4000$~K are
displayed in Fig.~\ref{f_fe}. 
One can observe that the magnetic moment $M_\mathrm{Fe}$, the
entropy $S$, and the total energy $E_\mathrm{tot}$ are monotonic
functions of the external field $B_\mathrm{ext}$ throughout the
studied range.
However, the free energy $F = E_\mathrm{tot} - TS$ exhibits a clear
minimum at $B_\mathrm{ext} = 0$, which represents a necessary
condition for an internally consistent theory at finite
temperatures.
Note that the total energy $E_\mathrm{tot}$ is expressed only
in terms of quantities discussed in Section~\ref{ss_ede}, i.e.,
it is fully independent of the treatment of the electronic entropy
$S$ described in Section~\ref{ss_entr}.

\section{Conclusion\label{s_concl}}

We have revised the evaluation of physical quantities in
finite-temperature Green's-function techniques with particular
attention paid to the electronic entropy.
We have shown that usual obstacles encountered in entropy
calculations (branch cuts or ambiguity of complex logarithm) can
be removed completely, which leads to a simple final expression
without an explicit contour integration.
The final result can be implemented both in semiempirical TB
schemes as well as in \emph{ab initio} Green's function approaches
based on the KKR or the LMTO methods, optionally also with the
CPA for chemically disordered systems.
The finite numerical accuracy of the developed formalism, which is
due to a standard auxiliary Taylor expansion of the resolvent,
seems to be well under control.
The electronic entropy thus need not be avoided in Green's function
techniques, but it should rather be employed directly for reliable
computations of other thermodynamic potentials. 

\begin{acknowledgments}
This work was supported financially by the Czech Science
Foundation (Grant No.\ 18-07172S).
\end{acknowledgments}

%\bibliography{b_}

\providecommand{\noopsort}[1]{}\providecommand{\singleletter}[1]{#1}%

\end{document}